\documentstyle[12pt]{article}
\let\tilde=\widetilde

\def\e{{\rm e}}
\def\phi{\varphi}
\def\eps{\varepsilon}

\def\l{\lambda}
\def\L{\Lambda}

\def\dd{\partial}

\def\one#1{#1^{\raise5pt\hbox{$\scriptstyle\!\!\!\!1$}}\,{}}
\def\two#1{#1^{\raise5pt\hbox{$\scriptstyle\!\!\!\!2$}}\,{}}
\def\three#1{#1^{\raise5pt\hbox{$\scriptstyle\!\!\!\!3$}}\,{}}
\def\ket#1{\left|#1\right>}
\def\bra#1{\left<#1\right|}

\def\abs#1{\left|#1\right|}
\def\id{\hbox{{1}\kern-.25em\hbox{\rm l}}}

\def\comment#1{}

\def\beq{\begin{equation}}
\def\eeq{\end{equation}}
\def\be{\begin{displaymath}}
\def\ee{\end{displaymath}}
\def\bea{\begin{eqnarray}}
\def\eea{\end{eqnarray}}
\def\beas{\begin{eqnarray*}}
\def\eeas{\end{eqnarray*}}
\def\bds{\begin{description}}
\def\eds{\end{description}}
\def\bmat{\left(\begin{array}}
\def\emat{\end{array}\right)}

\newtheorem{theo}{Theorem}

\def\half{\frac{1}{2}}

\def\Ref#1{(\ref{#1})}
\def\?{(?)\marginpar{|?}}
\def\add#1{{$\lceil$\sf #1$\rfloor$}\marginpar{{$\bullet$}}}
\newfont{\frak}{eufm10 scaled\magstep1}
\newfont{\bbd}{msbm10 scaled\magstep1} %% if mbsm font not available use
\def\C{\hbox{\bbd C}}                  %% \def\C{{\bf C}}
\def\eg{\hbox{\frak e}}
\def\fg{\hbox{\frak f}}
\def\hg{\hbox{\frak h}}

\def\CC{{\cal C}}
\def\M{{\cal M}}
\def\PP{{\cal P}}
\def\U{{\cal U}}
\def\V{{\cal V}}
\def\W{{\cal W}}
\def\E{{\cal E}}
\def\F{{\cal F}}
\def\H{{\cal H}}
\def\A{{\cal A}}
\def\X{{\cal X}}
\def\Y{{\cal Y}}
\def\pair#1#2{\left<#1\mid{#2}\right>}
\def\mean#1{\bra0#1\ket0}

\def\teps{\eps^\prime}
\def\tphi{\phi^\prime}
\def\teta{\eta^\prime}
\def\pmatrix#1{\left(\begin{array}{cc}#1\end{array}\right)}
\def\fpint{\frac{1}{2\pi i}\int_{\Gamma}}
\renewcommand{\theequation}{\thesection.\arabic{equation}}
\newcounter{subequation}[equation]
\makeatletter
 
\expandafter\let\expandafter
\reset@font\csname reset@font\endcsname
 
\def\subeqnarray{\arraycolsep1pt
    \def\@eqnnum\stepcounter##1{\stepcounter{subequation}%
        {\reset@font\rm(\theequation\alph{subequation})}}
\jot5mm     \eqnarray}

\makeatother
\newcommand{\newsection}[1]{
\vspace{8mm}
\addtocounter{section}{1}
\setcounter{equation}{0}
{\noindent\indent\large\bf\thesection.  #1}
\nopagebreak
\medskip
\nopagebreak}
\begin{document}
\begin{flushright}
 \bf PDMI 10/1997 \\
\bf solv-int/9708007
\end{flushright}
\vskip 1cm
\begin{center}
\LARGE\bf Generating function of correlators \\
in the $sl_2$ Gaudin model
\end{center}
\vskip 1cm
\begin{center}
E.K.~Sklyanin
\vskip 0.7cm
Steklov Mathematical Institute at St.~Petersburg, \\
Fontanka 27, St.~Petersburg 191011, Russia
\footnote{E-mail: {\tt sklyanin@pdmi.ras.ru}}
\end{center}
\vskip 1cm
\begin{center}
  June 1997
\end{center}
\vskip 1cm
\begin{center}
\bf Abstract
\end{center}
For the $sl_2$ Gaudin model (degenerated quantum integrable XXX spin chain)
an exponential generating function of correlators is calculated explicitely.
The calculation relies on the Gauss decomposition for the $SL_2$ loop group.
From the generating function a new explicit expression for the correlators
is derived from which the determinant formulas for the norms of Bethe
eigenfunctions due to Richardson and Gaudin are obtained.

\newpage
%%%%%%%%%%%%%%%%%%%%%%%%%%%%%%%%%%%%%%%%%%%%%%%%%%%%%%%%%%%%%%%%%%%%%%%%%%%%%%
\newsection{Introduction}

The most challenging problem of the theory of quantum integrable systems,
after describing the spectrum and eigenfunctions, is finding a reasonably
efficient expression for the correlators. Speaking about the systems lying
in the realm of Bethe ansatz (R-matrix, quantum inverse scattering) method,
two complementary approaches are presently known. One of them is based on
studying analyticity and bootstrap equations for factorized $S$-matrices
and formfactors of infinite-volume quantum field models and is useful in
the study of short-distance behaviour of correlators (see the review papers
\cite{Smirnov,JM}). Another one, summarized in the book \cite{KBI},
is, on the contrary,  efficient
in describing the long-distance asymptotics and is based on
meticulous study of the structure of Bethe eigenfunctions in the finite
volume.

Whereas the mathematical structures involved with the former approach
are now recognized quite satisfactorily (quantum Knizhnik-Zamolodchikov
equation, vertex operators, representations of quantum groups) which
has resulted in the rapid progress
in this subject, this is not the case for the latter approach.
After {\it tour de force} of 1982 by V.~Korepin \cite{Kor82},
who had managed to overcome
the enormous combinatorial difficulties in handling the cumbersome
expressions for correlators produced via algebraic Bethe ansatz, very little
progress has been achieved in mathematical understanding of his technique.
One should mention \cite{MS96} where Drinfeld's twists
were applied to simplify Korepin's calculations,
and \cite{RV} where an alternative derivation
of the determinant formula for the Bethe eigenfunctions based on the
semiclassical asymptotics of the solutions to the
Knizhnik-Zamolodchikov equation was proposed.

The present paper has grown from an attempt to understand and to simplify
Korepin's derivations using a simple toy example. As such, I have
chosen the $sl_2$ Gaudin model \cite{Gau67,Gau76,Gau83} (a degenerated case
of integrable XXX spin chain). The choice was motivated by the fact that
Gaudin model had already played this role once. The determinant formula
for the Bethe eigenfunctions of Gaudin model obtained by Gaudin,
who, in turn, had benefited from an earlier paper by Richardson \cite{Ri65},
allowed to Gaudin, McCoy and Wu to conjecture in \cite{GMW}
the similar formula for the XXX spin chain which was finally proved by
Korepin \cite{Kor82}.

The intimidating combinatorial complexity of the algebraic expressions for
correlators resulting from algebraic Bethe ansatz makes it natural to
take advantage of the common combinatorial wisdom which says that
using generating functions can help to reduce the complexity drastically.
As shown in the present paper, this simple idea works quite effectively
in case of the Gaudin model. A simple explicit expression for the exponential
generating function of correlators is obtained from the Gauss decomposition
for $SL(2)$ loop algebra. From the generating function the formulas for
particular correlators are derived quite easily, including the
Richardson-Gaudin determinant formula for the norm of Bethe eigenfunction.

The paper is organised as follows.
In Section 2 we introduce the $sl_2$ Gaudin model and present the known
results for its eigenfunctions, in particular, the Richardson-Gaudin
fomula. In Section 3, we describe a new formula,
called {\it $\L$-representation} for correlators which we prove
in the next, 4th section using the generating function technique.
The idea of the proof is explained first on the elementary finite-dimensional
example of $sl_2$ Lie algebra. The $\L$-representation is used in Section
5 to derive the Richardson-Gaudin determinant formula. In the concluding,
6th section we discuss the obtained results and the prospects of their
generalisation.

This work has been started during my stay at the Department of Mathematical
Sciences, the University of Tokyo, when I was preparing
a course of lectures for graduate students on Gaudin model
and was continued in Research Institute for Mathematical Studies,
Kyoto University. I am grateful
to UoT. and RIMS for support and hospitality.
%I wish also to express my gratitude
%to V.E.~Korepin, A.N.~Kirillov and V.O.~Tarasov for their interest in the
%work and useful discussions.

%%%%%%%%%%%%%%%%%%%%%%%%%%%%%%%%%%%%%%%%%%%%%%%%%%%%%%%%%%%%%%%%%%%%%%%%%%%%%
\newsection{Description of the model}

The aim of this section is mainly to fix the notation.
The results given below are taken from from \cite{Gau83,Skl24}.

Let $\eg$, $\fg$, $\hg$ be the generators of the Lie algebra $sl_2$:
\beq
 [\eg,\fg]=\hg \qquad [\hg,\eg]=2\eg \qquad [\hg,\fg]=-2\fg,
\label{eq:comehf}
\eeq
and let a highest weight module labelled with the complex number
(highest weight) $\l$ be
generated by the vacuum (highest vector) $\ket0$ and the relations
\beq
 \eg\ket0=0, \qquad \hg\ket0=\l\ket0.
\eeq

We shall need also the dual module generated by the dual vacuum $\bra0$
and relations
\beq
 \bra0\fg=0, \qquad \bra0\hg=\l\bra0, \qquad \pair00=1.
\eeq

The corresponding value of the Casimir operator
\beq
 2K=\eg\fg+\fg\eg+\half \hg^2
\label{eq:defKsl2}
\eeq
is then $2K=\l(\l+2)/2$.

The commuting Hamiltonians $\Xi_j$ of the Gaudin model are defined in the 
tensor product of $D$ h.w.\ modules (we label the corresponding generators with
the index $j$ and put $\ket0:=\ket0_1\otimes\cdots\otimes\ket0_D$) and are
given by the expression
\beq
 \Xi_j=\sum_{k\neq j}\frac{\eg_j\fg_k+\fg_j\eg_k+\half\hg_j\hg_k}{z_j-z_k}  ,
\qquad j=1,\ldots D
\label{eq:def-Xi}
\eeq
where $z_j$ are complex constants.
It is more convenient, however, to express $\Xi_j$ in terms of a generating
function. To this end, consider the one-parameter operator families
\beq
 E(u)=\sum_{j=1}^D\frac{\eg_j}{u-z_j}, \qquad
 F(u)=\sum_{j=1}^D\frac{\fg_j}{u-z_j}, \qquad
 H(u)=\sum_{j=1}^D\frac{\hg_j}{u-z_j}, \qquad
\label{eq:EFH-efh}
\eeq
and define the generating function $t(u)$ as the result of replacing
$\eg$, $\fg$, $\hg$ in the Casimir \Ref{eq:defKsl2} respectively with
$E(u)$, $F(u)$ and $H(u)$:
\beq
 2t(u)=E(u)F(u)+F(u)E(u)+\half H^2(u)
\eeq

Using \Ref{eq:EFH-efh} and \Ref{eq:def-Xi} one observes that
\beq
 t(u)=\sum_{j=1}^D \frac{\Xi_j}{u-z_j}+
\frac{\frac14\l_j(\l_j+2)}{(u-z_j)^2}
\eeq

The fundamental property of the operators \Ref{eq:EFH-efh} is that they form
the highest-weight module over the infinite-dimensional Lie algebra
\cite{Gau83,Skl24}
\begin{subeqnarray}
{[}E(u),E(v){]}&=&{[}F(u),F(v){]}={[}H(u),H(v){]}=0, \\
{[}E(u),F(v){]}&=&-\frac{H(u)-H(v)}{u-v}, \\
{[}H(u),E(v){]}&=&-2\frac{E(u)-E(v)}{u-v}, \\
{[}H(u),F(v){]}&=&2\frac{F(u)-F(v)}{u-v},
\label{eq:comEFH}
\end{subeqnarray}
characterised by the vacuum $\ket0$,
\beq
 E(u)\ket0=0, \qquad H(u)\ket0=\l(u)\ket0,
\label{eq:EH-vac}
\eeq
dual vacuum $\bra0$
\beq
 \bra0F(u)=0, \qquad \bra0H(u)=\l(u)\bra0, \qquad
\pair00=1,
\label{eq:EH-dvac}
\eeq
and the scalar function (highest weight) $\l(u)$
\beq
 \l(u)=\sum_{j=1}^D\frac{\l_j}{u-z_j}\,.
\label{eq:def-lambda}
\eeq

A direct consequence of the relations \Ref{eq:comEFH} is,
in particular, the commutativity of $t(u)$
\beq
  [t(u),t(v)]=0
\eeq
from which the commutativity of the hamiltonians $\Xi_j$ follows immediately.

The commutation relations \Ref{eq:comEFH} can be also written
down compactly in the so-called {\it $r$-matrix} form
\cite{Gau83,Sem,Skl24} which we, however, shall not use here.

The theory of Gaudin model is expected to give solution to two main problems.
The first one is to determine the eigenvalues and eigenvectors of the
commuting Hamiltonians $\Xi_j$ \Ref{eq:def-Xi}.

The solution is given in terms
of {\it Bethe vectors} which are defined for any finite set $\V$ of complex
numbers as
\beq
 \ket\V:=\prod_{v\in\V}F(v)\ket0, \qquad \V\subset\C, \qquad \abs{\V}<\infty.
\eeq

\begin{theo} \cite{Gau67,Gau83,Skl24}
The vector $\ket\V$ is a joint eigenvector of commuting operators $\Xi_j$ or,
equivalently, of $t(u)$ if and only if the parameters $v\in\V$ satisfy the
{\em Bethe equations}
\beq
 \l(v)=\sum_{v'\neq v}\frac{2}{v-v'}, \qquad \forall v\in\V.
\label{eq:bethe}
\eeq

The corresponding eigenvalue $\tau(u)$ of $t(u)$ is then
\beq
 \tau(u)=\frac14\tilde\l^2(u)-\frac12\partial_u\tilde\l(u)
\eeq
where
\beq
 \tilde\l(u):=\l(u)-\sum_{v\in\V}\frac{2}{u-v}.
\eeq
\label{th:Bethe-eqs}
\end{theo}

The spectrum of the model having been determined, the next fundamental
problem is to
calculate the correlators. The correlators which we are going to study are 
labelled by three finite sets $\U,\V,\W\subset\C$ and are defined as
\beq
 C(\U,\W,\V):=\bra\U\prod_{w\in\W}H(w)\ket\V
\eeq
where $\bra\U$ is the dual Bethe vector
\beq
 \bra\U:=\bra0\prod_{u\in\U}E(u).
\eeq

In principle, one could calculate $C(\U,\W,\V)$ moving the operators $H(w)$
and $E(u)$ to the right through $F(v)$ with the use of the formulas
\begin{subeqnarray}
 F(v)\ket\V&=&\ket{\V\cup\{v\}}, \\
 H(w)\ket\V&=&\left(\l(w)-\sum_{v\in\V}\frac{2}{w-v}\right)\ket\V
+\sum_{v\in\V}\frac{2}{w-v}\ket{\V\setminus\{v\}\cup\{w\}}, \\
 E(u)\ket\V&=&\sum_{v\in\V}-\frac{1}{u-v}
\left(\l(u)-\l(v)-\sum_{v'\neq v}
2\left(\frac{1}{u-v'}-\frac{1}{v-v'}\right)\right)
\ket{\V\setminus\{v\}} \nonumber \\
&&+\sum_{\scriptstyle v,v'\in\V \atop \scriptstyle v\neq v'}
-\frac{2}{(u-v)(u-v')}
\ket{\V\setminus\{v,v'\}\cup\{u\}}
\label{eq:Korepin-formulas}
\end{subeqnarray}
which follow from the relations \Ref{eq:comEFH} and 
\Ref{eq:EH-vac}. For example
\beq
 C(\{u\},\emptyset,\{v\})=-\frac{\l(u)-\l(v)}{u-v}, \qquad u\neq v.
\label{eq:C(u,0,v)}
\eeq

This strategy was adopted in \cite{Kor82} where the formulas
similar to \Ref{eq:Korepin-formulas} were used in a more complicated case of
XXX spin chain. Such straightforward approach leads, however, to extremely
tedious calculations, and it is the aim of the present paper to simplify
the derivation using the method of generating functions. Still,
one simple observation can be made right now: the correlator $C(\U,\W,\V)$
is a polynomial in $\l(u)$, $\l(v)$, $\l(w)$ with the coeffitients rational
in $u\in\U$, $v\in\V$, $w\in\W$.

It is important to stress, following \cite{Kor82}, 
that for the derivation of the formulas for 
$C(\U,\W,\V)$ the concrete form \Ref{eq:def-lambda} of the function
$\l(u)$ is not important, it can be considered as a formal parameter.
All what matters is analyticity of $\l(u)$ at the points of
$\U\cup\V\cup\W$ and
the relations \Ref{eq:comEFH}, \Ref{eq:EH-vac}
and \Ref{eq:EH-dvac}.

We conclude this list of known results with an important determinant formula
for the norm of the Bethe eigenfunction. 
\begin{theo}
Let $\V=\{v_1,\ldots,v_N\}$ and $v_j$ satisfy the Bethe equations
\Ref{eq:bethe}. Then
\beq
 \pair\V\V=(-1)^N\det\M
\label{eq:norm}
\eeq
where the matrix $\M$ is given by
\begin{subeqnarray}
 \M_{jj}&=&\dd_{v_j}\l(v_j)+\sum_{k\neq j}\frac{2}{(v_j-v_k)^2},  \\
 \M_{jk}&=&-\frac{2}{(v_j-v_k)^2}, \qquad j\neq k
\end{subeqnarray}
\label{th:norm}
\end{theo}

The formula \Ref{eq:norm} was derived first by 
Richardson \cite{Ri65} for a model which can now be recognized as a very
degenerate case of Gaudin model. In \cite{Gau76} Gaudin mentioned that
the norms for Gaudin model are given by the same formula
but had not explained how to generalise Richardson's derivation.
To make up the deficiency we give a complete proof of \Ref{eq:norm}
in section 5.

%%%%%%%%%%%%%%%%%%%%%%%%%%%%%%%%%%%%%%%%%%%%%%%%%%%%%%%%%%%%%%%%%%%%%%%%%%%
\newsection{$\L$-representation for correlators}

In this section we give a new explicit formula, which we call 
$\L$-representation, for the correlator $C(\U,\W,\V)$. Throughout this
paper we suppose, unless otherwise stated, that the sets $\U$, $\V$ and $\W$
do not pairwise intersect.

From \Ref{eq:Korepin-formulas} it follows immediately that $C(\U,\W,\V)$
is a polynomial in
$\l(x)$, $x\in\U\cup\V\cup\W$ with the coefficients rational in $x$.
However, the structure of $C(\U,\W,\V)$ can be simplified
after introducing a proper notation.

For any finite set $\X\subset\C$ define the function
\beq
 \L(\X):=\sum_{x\in\X}
\frac{\l(x)}{\prod\limits_{y\neq x}(x-y)}.
\label{eq:def-L}
\eeq

For example,
\beq
 \L(\{x\})=\l(x), \qquad
 \L(\{x_1,x_2\})=\frac{\l(x_1)-\l(x_2)}{x_1-x_2}.
\label{eq:L-ex}
\eeq

Note that, supposing $\l(u)$ to be analytic inside a contour $\Gamma$
circumscribing counterclockwise the set $\X$, one can rewrite \Ref{eq:def-L}
as the contour integral
\beq
   \L(\X)=\fpint\frac{\l(z)dz}{\prod\limits_{x\in\X}(z-x)},
   \qquad \X\subset\C_+\,.
\label{eq:def-L-int}
\eeq

Our aim is to express $C(\U,\W,\V)$ as a polynomial in $\L$.

Let $\abs{\cdot}$ be the cardinality of a set, and suppose
\beq
 \abs\U=\abs\V=N, \qquad \abs\W=M
\eeq
(obviously, $C(\U,\W,\V)$=0 if $\abs{\U}\neq\abs{\V}$).
Let us define
a {\it coordinated partition}
$\PP$ of the sets $\U$, $\V$ and $\W$
as a set of triplets $P=(\U_P,\V_P,\W_P)$, $\U\supset\U_P\neq\emptyset$,
$\V\supset\V_P\neq\emptyset$, $\W\supset\W_P$ (note that $\W_P$, in contrast 
with $\U_P$ and $\V_P$, is allowed to be empty)
such that
\be
  \forall P,P'\in\PP:\qquad \abs{\U_P}=\abs{\V_P}>0, \qquad \abs{\W_P}\geq0,
\ee
\be
P\neq P'\qquad \Longrightarrow\qquad
 \U_P\cap\U_{P'}=\emptyset, \quad \V_P\cap\V_{P'}=\emptyset,
\quad \W_P\cap\W_{P'}=\emptyset,
\ee
\be
 \bigcup_{P\in\PP}\U_P=\U, \qquad \bigcup_{P\in\PP}\V_P=\V, 
\qquad \bigcup_{P\in\PP}\W_P\subset\W,
\ee
\be
 \sum_{P\in\PP}\abs{\U_P}=\abs{\U}=\sum_{P\in\PP}\abs{\V_P}=\abs\V=N, \qquad
 \sum_{P\in\PP}\abs{\W_P}\leq\abs\W=M.
\ee

\begin{theo}
The expression for the correlator $C(\U,\W,\V)$
is given by the formula
\bea\lefteqn{
C(\U,\W,\V)}\nonumber \\
&&=(-1)^N\sum_{\PP}\left(\prod_{P\in\PP}
 \abs{\U_P}!\,(\abs{\U_P}-1)!\,(2\abs{\U_P})^{\abs{\W_P}}\,
\L(\U_P\cup\V_P\cup\W_P)\right) \nonumber \\
&&\hphantom{(-1)^N\sum}\times\left(\prod_{w\in\overline\W_\PP}\l(w)\right)
\label{eq:L-repr-gen}
\eea
where 
\beq
\overline\W_\PP:=\W\setminus\bigcup_{P\in\PP}\W_P.
\eeq
\label{th:lambda-repr}
\end{theo}

For example,
\be
 C(\{u\},\{w\},\{v\})=-2\L(\{u,w,v\})-\L(\{u,v\})\L(\{w\}).
\ee

It is remarkable that $C(\U,\W,\V)$ is a polynomial  in $\L$ with
{\em integer} coefficients, that
is all the rationality is hidden in the definition \Ref{eq:def-L} of $\L$.

To describe the norms of Bethe eigenfunctions we shall need a particular
case of \Ref{eq:L-repr-gen} when $\W=\emptyset$:
\beq
 \pair\U\V=(-1)^N\sum_{\PP}\prod_{P\in\PP}
 \abs{\U_P}!\,(\abs{\U_P}-1)!\,\L(\U_P\cup\V_P)
\label{eq:L-repr-norm}
\eeq
where $\PP$ runs over the
sets of pairs $P=(\U_P,\V_P)$, $\U\supset\U_P\neq\emptyset$,
$\V\supset\V_P\neq\emptyset$,
such that
\be
  \forall P\in\PP:\qquad \abs{\U_P}=\abs{\V_P},
\ee
\be
 P,P'\in\PP,\quad P\neq P'\qquad \Longrightarrow\qquad
 \U_P\cap\U_{P'}=\emptyset, \quad \V_P\cap\V_{P'}=\emptyset,
\ee
\be
 \bigcup_{P\in\PP}\U_P=\U, \qquad \bigcup_{P\in\PP}\V_P=\V,
\ee
\be
 \sum_{P\in\PP}\abs{\U_P}=\abs\U=\sum_{P\in\PP}\abs{\V_P}=\abs\V=N.
\ee

For example,
\begin{subeqnarray}
\pair00&=&1, \qquad
\pair uv=-\L(\{u,v\}), \\
\pair{u_1u_2}{v_1v_2}&=&
\L(\{u_1,v_1\})\L(\{u_2,v_2\})+\L(\{u_1,v_2\})\L(\{u_2,v_1\})  \nonumber \\
&&+2\L(\{u_1,u_2,v_1,v_2\}).
\label{eq:L-repr-simp}
\end{subeqnarray}

Using the integral formula \Ref{eq:def-L-int} for $\L(\X)$ it is easy
to extend by analyticity the formula \Ref{eq:L-repr-gen} for
$C(\U,\W,\V)$ to the case when the sets $\U$, $\V$, $\W$ intersect or
even have multiple points (being thus not sets but divisors). We leave this
exercise to the reader.

%%%%%%%%%%%%%%%%%%%%%%%%%%%%%%%%%%%%%%%%%%%%%%%%%%%%%%%%%%%%%%%%%%%%%%%%%%%%%
\newsection{Proof of the Theorem \ref{th:lambda-repr}}

As mentioned in the Introduction, we are going to circumvent the combinatorial
difficulties of the direct approach based on the identities
\Ref{eq:Korepin-formulas} by calculating a generating function of
correlators $C(\U,\W,\V)$.

Let us illustrate the idea on an
elementary finite-dimensional example of the $sl_2$ Lie algebra 
\Ref{eq:comehf}. The analog of $C(\U,\W,\V)$ is the expression
$c(k,j):=\mean{\eg^k\hg^j\fg^k}$ which can easily be estimated as
\beq
c(k,j)=(-k)_k(-\l)_k(\l-2k)^j=(-1)^k\,k!\,(-\l)_k(\l-2k)^j
\label{eq:ckj}
\eeq
where the Pochhammer symbol is defined as
\beq
 (a)_k:=a(a+1)\ldots(a+k-1).
\eeq

On the other hand, let us introduce the group elements
\beq
 \E_\eps:=\e^{\eps\eg}, \qquad
 \H_\eta:=\e^{\eta\hg}, \qquad
 \F_\phi:=\e^{\phi\fg}, \qquad
\eeq
parametrized by coordinates $\eps$, $\eta$, $\phi$ and 
consider the exponential generating function
\beq
 \mean{\E_\eps \H_\eta \F_\phi}=\sum_{k=0}^\infty\sum_{j=0}^\infty
\frac{\eps^k\eta^j\phi^k}{k!^2\,j!}c(k,j).
\eeq

Notice now that if we manage to change the order of factors 
\beq
 \E_\eps \H_\eta \F_\phi = \F_{\tphi}\H_{\teta}\E_{\teps},
\label{eq:gauss-id}
\eeq
then the generating function is easily calculated:
\beq
 \mean{\F_{\tphi}\H_{\teta}\E_{\teps}}=\mean{\H_{\teta}}=\e^{\teta\l}.
\eeq

To establish the identity \Ref{eq:gauss-id} it is sufficient to consider the
fundamental (2-dimensional) representation of Lie algebra $sl_2$
\beq
 \eg=\pmatrix{0 & 1 \\ 0 & 0}, \qquad
 \hg=\pmatrix{1 & 0 \\ 0 & -1}, \qquad
 \fg=\pmatrix{0 & 0 \\ 1 & 0}.
\eeq
and the corresponding Lie group $SL_2$
\beq
 \E_\eps=\pmatrix{1 & \eps \\ 0 & 1}, \qquad
 \H_\eta=\pmatrix{\e^{\eta} & 0 \\ 0 & \e^{-\eta}}, \qquad
 \F_\phi=\pmatrix{1 & 0 \\ \phi & 1}. \qquad
\eeq

The identity \Ref{eq:gauss-id} expresses then equivalence of left and
right Gauss decompositions in $SL_2$.
Multiplying $2\times2$ matrices we find from \Ref{eq:gauss-id} the values
of the primed parameters:
\beq
 \e^{\teta}=\e^{\eta}+\eps\phi \e^{-\eta}, \qquad
 \teps=\frac{\eps \e^{-\eta}}{\e^{\eta}+\eps\phi \e^{-\eta}}, \qquad
 \tphi=\frac{\phi \e^{-\eta}}{\e^{\eta}+\eps\phi \e^{-\eta}},
\eeq
whence
\beq
 \mean{\E_\eps \H_\eta \F_\phi}=(\e^{\eta}+\eps\phi \e^{-\eta})^\l
=\e^{\l\eta}(1+\eps\phi \e^{-2\eta})^\l.
\label{eq:meanEHF}
\eeq

Applying the binomial expansion formula 
\be
 (1-z)^{-a}=\sum_{k=0}^\infty \frac{(a)_k}{k!}\,z^k
\ee
to \Ref{eq:meanEHF} one recovers the result \Ref{eq:ckj}.

%To make the argument more rigorous, one should understand all the above
%equalities in the sense of formal series in $\eps$, $\eta$, $\phi$, since
%for generic $\l$ the representation of $sl_2$ is non-integrable.

The same Gauss decomposition trick works also in the case of
the infinite-dimensional Lie algebra \Ref{eq:comEFH}.

Let $\C_+$ be a simply connected bounded open domain on the complex plain
having smooth boundary
$\Gamma$ which is homeomorphic to a circle $S^1$. Let $\C_-$ be the complement
domain: $\C_-=\C\setminus(\C_+\cup\Gamma)$. Suppose that the contour $\Gamma$
is oriented counterclockwise.
Suppose also that $\U,\V,\W\subset\C_+$ and
that $\l(z)$ is analytical in $\C_+$.
%\add{comment: not very serious
%restrictions, since all the construction is just a compact way of
%manipulating formal series}.

Let $\eps(x)$, $\phi(x)$, $\eta(x)$ be some smooth functions on $\Gamma$.
Define the generators of the Lie algebra $\A$ as
\beq
 E_\eps=\fpint\eps(x)E(x), \quad
 H_\eta=\fpint\eta(x)H(x), \quad
 F_\phi=\fpint\phi(x)F(x), \quad
\label{eq:def-EeHhFf}
\eeq

To describe the commutation relations between them we need to introduce the
decomposition
\be
 \psi(x)=\psi_+(x)+\psi_-(x)
\ee
of any function $\psi(x)$ on the contour $\Gamma$ into the parts
$\psi_\pm(z)$ analytical respectively in $z\in\C_\pm$. It is supposed that
$\psi_-(z)\rightarrow0$ as $z\rightarrow\infty$.
The projections are given by the Cauchy integral
\be
 \psi_\pm(z):=\pm\fpint\frac{\psi(x)dx}{x-z}, \qquad
x\in\Gamma, \quad z\in\C_\pm\,.
\ee

Note the formula for the 
Hilbert transform  defined as the singular integral
\be
 \fpint\frac{\psi(y)dy}{x-y}=\frac{\psi_+(x)-\psi_-(x)}{2},
\qquad x,y\in\Gamma
\ee
regularized in the sense of the principal value.

Now we can write down the commutation relations for the generators
\Ref{eq:def-EeHhFf} as
\begin{subeqnarray}
 {[}E_{\eps^{(1)}},E_{\eps^{(2)}}{]}&=&
 {[}H_{\eta^{(1)}},H_{\eta^{(2)}}{]}=
 {[}F_{\phi^{(1)}},F_{\phi^{(2)}}{]}=0,\\
 {[}E_\eps,F_\phi{]}&=&H_{\eps_+\phi_+-\eps_-\phi_-}\,, \\
 {[}H_\eta,E_\eps{]}&=&2E_{\eta_+\eps_+-\eta_-\eps_-}\,, \\
 {[}H_\eta,F_\phi{]}&=&-2F_{\eta_+\phi_+-\eta_-\phi_-}\,,
\end{subeqnarray}

The above formulas suggest the decomposition, see \cite{Sem},
of $\A$ into two mutually commutative subalgebras $\A=\A_++\A_-$ generated,
respectively, by $E_{\eps_\pm}$, $H_{\eta_\pm}$, $F_{\phi_\pm}$:
\be
E_\eps=E_{\eps_+}+E_{\eps_-}, \qquad
H_\eta=H_{\eta_+}+H_{\eta_-}, \qquad
F_\phi=F_{\phi_+}+F_{\phi_-}.
\ee

The commutation relations for the subalgebras $\A_\pm$ are those for the
$sl_2$ loop Lie algebra
\begin{subeqnarray}
\kern-1cm {[}E_{\eps_\pm^{(1)}},E_{\eps^{(2)}_\pm}{]}&=&
 {[}H_{\eta_\pm^{(1)}},H_{\eta^{(2)}_\pm}{]}=
 {[}F_{\phi_\pm^{(1)}},F_{\phi^{(2)}_\pm}{]}=0, \\
\kern-1cm {[}E_{\eps_\pm},F_{\phi_\pm}{]}&=&\pm H_{\eps_\pm\phi_\pm}, \quad
 {[}H_{\eta_\pm},E_{\eps_\pm}{]}=\pm2E_{\eta_\pm\eps_\pm}, \quad
 {[}H_{\eta_\pm},F_{\phi_\pm}{]}=\mp2F_{\eta_\pm\phi_\pm},
\end{subeqnarray}
and allow the fundamental representation
as $2\times2$ matrix-valued functions of $x\in\Gamma$
\be
 E_{\eps_\pm}=\pm\eps_\pm(x)\pmatrix{0 & 1 \\ 0 & 0}, \quad
 H_{\eta_\pm}=\pm\eta_\pm(x)\pmatrix{1 & 0 \\ 0 & -1}, \quad
 F_{\phi_\pm}=\pm\phi_\pm(x)\pmatrix{0 & 0 \\ 1 & 0}.
\ee
with the pointwise commutator.

Now everything is ready to reproduce the same Gauss decomposition trick
which we saw working in the finite-dimensional case.

Introduce the group elements
\be
  \E_{\eps_\pm}:=\exp E_{\eps_\pm}, \qquad
  \H_{\eta_\pm}:=\exp H_{\eta_\pm}, \qquad
  \F_{\phi_\pm}:=\exp F_{\phi_\pm},
\ee
having the fundamental representation
\be
 \E_{\eps_\pm}=\pmatrix{1 & \pm\eps_\pm(x) \\ 0 & 1}, \quad
 \H_{\eta_\pm}=
 \pmatrix{\e^{\pm\eta_\pm(x)} & 0 \\  0 & \e^{\mp\eta_\pm(x)}}, \quad
 \F_{\phi_\pm}=\pmatrix{1 & 0 \\ \pm\phi_\pm(x) & 1}.
\ee
%\add{argument $x$}

The identity
\beq
  \E_{\eps_\pm}\H_{\eta_\pm}\F_{\phi_\pm}=
  \F_{\tphi_\pm}\H_{\teta_\pm}\E_{\teps_\pm}
\eeq
is established by the same calculation as in the finite-dimensional case
\beq
 \e^{\pm\teta_\pm}=\e^{\pm\eta_\pm}+\eps_\pm\phi_\pm \e^{\mp\eta_\pm},
\eeq
\beq
 \teps_\pm=\frac{\eps_\pm \e^{\mp\eta_\pm}}{\e^{\pm\eta_\pm}
          +\eps_\pm\phi_\pm \e^{\mp\eta_\pm}}, \qquad
 \tphi_\pm=\frac{\phi_\pm \e^{\mp\eta_\pm}}{\e^{\pm\eta_\pm}
          +\eps_\pm\phi_\pm \e^{\mp\eta_\pm}},
\eeq
one should rememeber only that now $\eps_\pm$, etc.\ depend on parameter
$x\in\Gamma$.

The formula for the generating function of correlators follows then
immediately:
\bea
  \left<\E_\eps\H_\eta\F_\phi\right>&=&
   \left<\E_{\eps_+}\H_{\eta_+}\F_{\phi_+}\right>
   \left<\E_{\eps_-}\H_{\eta_-}\F_{\phi_-}\right> \nonumber \\
   &=& \left<\F_{\tphi_+}\H_{\teta_+}\E_{\teps_+}\right>
       \left<\F_{\tphi_-}\H_{\teta_-}\E_{\teps_-}\right> \nonumber \\
   &=&\left<\H_{\teta_+}\right>\left<\H_{\teta_-}\right>
   =\exp\fpint\l(z)\teta(z)dz
\label{eq:corr-EHF}
\eea
where
\beq
\teta=\teta_++\teta_-, \qquad
 \teta_\pm=
     \pm\ln(\e^{\pm\eta_\pm}+\e^{\mp\eta_\pm}\eps_\pm\phi_\pm).
\eeq

Now we can prove Theorem \ref{th:lambda-repr}. Put
\beq
  \eps(z):=\sum_{u\in\U}\frac{\eps_u}{z-u}, \qquad
  \eta(z):=\sum_{w\in\W}\frac{\eta_w}{z-w}, \qquad
  \phi(z):=\sum_{v\in\V}\frac{\phi_v}{z-v},
\label{eq:def-ehf(z)}
\eeq
where $\eps_u,\eta_w,\phi_v\in\C$.
Then, since by assumption $\U\subset\C_+$,
\beq
 E_\eps=\exp\fpint\sum_{u\in\U}\frac{\eps_u E(z)dz}{z-u}=
\exp\sum_{u\in\U}\eps_u E(u),
\eeq
and similarly for $H_\eta$, $F_\phi$. The correlator $C(\U,\W,\V)$ is given
then by the coefficient at
\be
\left(\prod_{u\in\U}\eps_u\right)
\left(\prod_{v\in\V}\phi_v\right)
\left(\prod_{w\in\W}\eta_w\right)
\ee
in the expansion of the generating function \Ref{eq:corr-EHF} in powers of
$\eps_u$, $\eta_w$, $\phi_v$.

To perform the expansion, note first that
\beq
  \eps_+(z)=0, \qquad \eps_-(z)=\eps(z)
\eeq
and similarly for $\phi(z)$, $\eta(z)$. Respectively, $\teta_+=0$ and
\bea
\teta=\teta_-&=&-\ln(\e^{-\eta}+\eps\phi\e^{\eta})=
   \eta-\ln(1+\eps\phi\e^{2\eta}) \nonumber \\
&=&\eta-\sum_{k=1}^\infty\sum_{j=0}^\infty
\frac{(-1)^{k-1}}{k}\frac{(2k)^j}{j!}
\eps^k\phi^k\eta^j,
\label{eq:expand-teta}
\eea
where we omitted the argument $z$ of $\eps(z)$, $\eta(z)$, $\phi(z)$.
As it was explained above, we are interested only in the terms linear in 
$\eps_u$, $\phi_v$, $\eta_w$.
For example, the contribution of $\eps^k$ is
\beq
 k!\,\sum_{\scriptstyle\U'\subset\U\atop\scriptstyle \abs{\U'}=k}
\prod_{u\in\U'}\frac{\eps_u}{z-u}\,.
\eeq

Respectively, the relevant terms in $\teta(z)$ are
\bea\lefteqn{\kern-1.5cm
 \sum_{w\in\W}\frac{\eta_w}{z-w}
+\sum_{\scriptsize\U'\subset\U\atop\scriptsize \abs{\U'}=\abs{\V'}}
\sum_{\scriptsize\V'\subset\V\atop\scriptsize \V'\neq\emptyset}
\sum_{\W'\subset\W}
(-1)^{\abs{\U'}}\,\abs{\U'}!\,(\abs{\U'}-1)!\,(2\abs{\U'})^{\abs{\W'}}
}\nonumber \\
&&\times\prod_{u\in\U'}\frac{\eps_u}{z-u}
\prod_{v\in\V'}\frac{\phi_v}{z-v}
\prod_{w\in\W'}\frac{\eta_w}{z-w}\,.
\label{eq:relevant-teta}
\eea

Integrating the result with $\l(z)$ and using the equality \Ref{eq:def-L-int}
%\add{used analyticity of $\l(u)$!!!}
we obtain finally for the relevant contribution to
$(1/2\pi i)\int\l(z)\teta(z)dz$:
\bea
 \sum_{w\in\W}\eta_w\l(w)
&+&\sum_{\scriptsize\U'\subset\U\atop\scriptsize \abs{\U'}=\abs{\V'}}
\sum_{\scriptsize\V'\subset\V\atop\scriptsize \V'\neq\emptyset}
\sum_{\W'\subset\W}
(-1)^{\abs{\U'}}\,\abs{\U'}!\,(\abs{\U'}-1)!\,(2\abs{\U'})^{\abs{\W'}}
\nonumber \\
&&\times\left(\prod_{u\in\U'}\eps_u\right)
\left(\prod_{v\in\V'}\phi_v\right)
\left(\prod_{w\in\W'}\eta_w\right)
\L(\U'\cup\V'\cup\W')
\label{eq:Lteta}
\eea

Exponentiating \Ref{eq:Lteta} and retaining the term containing the factor
$\prod\eps_u\phi_v\eta_w$
we finally obtain \Ref{eq:L-repr-gen}.

%%%%%%%%%%%%%%%%%%%%%%%%%%%%%%%%%%%%%%%%%%%%%%%%%%%%%%%%%%%%%%%%%%%%%%%%%%%%%%
\newsection{Richardson's determinant formula}

For the rest of this paper we consider only the case $\W=\emptyset$.
To prove the Theorem \ref{th:norm}, we derive first a recurrence relation for
the correlators $C(\U,\emptyset,\V)=\pair{\U}{\V}$.

Note the following property of the function $\L$
\beq
 \L(\X\cup\Y)=\sum_{x\in\X}\sum_{y\in\Y}
\frac{\L(\{x,y\})}%
{\prod\limits_{x'\neq x}(x-x')
 \prod\limits_{y'\neq y}(y-y')}
\label{eq:expand-L}
\eeq
which holds for $\X\cap\Y=\emptyset$ and can be deduced easily from the
definition \Ref{eq:def-L} of $\L$.

Using (\ref{eq:expand-L}) one can rewrite the $\L$-representation
\Ref{eq:L-repr-norm} for $\pair{\U}{\V}$
as a polynomial in the two-point correlators
$\L(\{u,v\})$ with the coefficients rational in $u$, $v$.
Let us denote the polynomial in question as $\CC(\U,\V;\ell)$ considering
$\ell(u,v)\equiv\L(\{u,v\})$ as a functional parameter:
$$
   \pair{\U}{\V}=\CC(\U,\V;\L(\{u,v\})).
$$

Richardson \cite{Ri65} has found an alternative description of the function
$\CC(\U,\V;\ell)$.

\begin{theo}
The function $\CC(\U,\V;\ell)$ possesses the following properties:
\begin{enumerate}
 \item $\CC(\U,\V;\ell)$ is a polynomial in $\ell(u,v)$, $u\in\U$, $v\in\V$.
 \item Each of the $\ell(u,v)$ enters $\CC(\U,\V;\ell)$ linearly.
 \item The polynomial $\CC(\U,\V;\ell)$ has no free term, that is
\be
\CC(\U,\V;0)=0.
\ee
 \item Recurrence relation.
 Let $\tilde\U=\U\cup\{\tilde u\}$ and $\tilde\V=\V\cup\{\tilde v\}$.
Then the coefficient at $\ell(\tilde u,\tilde v)$ in
$\CC(\tilde\U,\tilde\V;\ell)$ is equal to $-\CC(\U,\V;\tilde\ell)$ where
\beq
 \tilde\ell(u,v):=\ell(u,v)+
 \frac{2}{(u-\tilde u)(v-\tilde v)}.
\label{eq:rec-rel-C}
\eeq
%\add{remark: nonsymmetric in $u$, $v$!}
 \item Initial condition
\be
   \CC(\{u\},\{v\};\ell)=\ell(u,v).
\label{eq:in-cond-C}
\ee
\end{enumerate}
\label{th:charact-FN}
\end{theo}

Obviously, the properties 1--5, if true, characterize the function
$\CC(\U,\V;\ell)$ uniquely.

Richardson himself has proved the above theorem by a straightforward
calculation for a very degenerate case of Gaudin model.
Our proof, valid for general Gaudin model,
is based on the $\L$-representation for $\pair{\U}{\V}$ and
is much simpler due to using the machinery of generating functions.

The statements 1, 2 and 3 of the Theorem follow directly from
the $\L$-representation \Ref{eq:L-repr-norm} and \Ref{eq:expand-L},
the statement 5 follows from
\Ref{eq:C(u,0,v)} and \Ref{eq:L-ex}. The remaining statement 4
(recurrence relation), in principle, can also be derived directly from
the $\L$-representation which leads, however, to lengthy combinatorial
calculations. As in case of the theorem \ref{th:lambda-repr}, using
generating functions again allows to simplify the derivation, .

Putting $\W=\emptyset$ in \Ref{eq:Lteta} one obtains the relevant
(that is linear in $\eps_u$, $\phi_v$) contribution to
$(1/2\pi i)\int\l(z)\teta(z)dz$:
\beq
\sum_{\scriptsize\U'\subset\U\atop\scriptsize \abs{\U'}=\abs{\V'}}
\sum_{\scriptsize\V'\subset\V\atop\scriptsize \V'\neq\emptyset}
(-1)^{\abs{\U'}}\,\abs{\U'}!\,(\abs{\U'}-1)!
\left(\prod_{u\in\U'}\eps_u\right)
\left(\prod_{v\in\V'}\phi_v\right)
\L(\U'\cup\V').
\label{eq:Lteta-uv-1}
\eeq

Let us find, first, how the substitution $\ell\mapsto\tilde\ell$
\Ref{eq:rec-rel-C} can be expressed in terms of the expansion
\Ref{eq:Lteta-uv-1}.
From \Ref{eq:expand-L} it follows that replacing all $\L(\{u,v\})$ with
\Ref{eq:rec-rel-C}
is equivalent to replacing all $\L(\U'\cup\V')$  in \Ref{eq:Lteta-uv-1}
with
\beq
  \L(\U'\cup\V')+\sum_{u\in\U'}\sum_{v\in\V'}
\frac{2}{(u-\tilde u)(v-\tilde v)}
\left(\prod_{u'\neq u}\frac{1}{u-u'}\right)
\left(\prod_{v'\neq v}\frac{1}{v-v'}\right).
\label{eq:replace-L}
\eeq

Substituting \Ref{eq:replace-L} into \Ref{eq:Lteta-uv-1} we conclude,
similarly to the derivation of \Ref{eq:relevant-teta}, that
the terms added to $(1/2\pi i)\int\l(z)\teta(z)dz$
coincide exactly with the expansion of
\be
-2\ln\left(1+\eps(\tilde u)\phi(\tilde v)\right)
\ee
where $\eps(z)$ and $\phi(z)$ are given by \Ref{eq:def-ehf(z)}.

Exponentiating, one obtains that the substitution $\ell\mapsto\tilde\ell$
is equivalent to multiplying the generating function
$\left<\E_\eps\F_\phi\right>$ by the factor
\beq
 \exp\left\{-2\ln\left(1+\eps(\tilde u)\phi(\tilde v)\right)\right\}
 =\left(1+\eps(\tilde u)\phi(\tilde v)\right)^{-2}.
\label{eq:(1+t)^{-2}}
\eeq

Using the binomial series
\be
  (1+t)^{-2}=\sum_{k=0}^\infty (-1)^k(k+1)t^k
\ee
one can expand the expression \Ref{eq:(1+t)^{-2}} in powers of $\eps_u$,
$\phi_v$ (see again the derivation of \Ref{eq:relevant-teta}
from \Ref{eq:expand-teta}) and retain the relevant terms obtaining finally
\beq
\sum_{\scriptsize\U'\subset\U\atop\scriptsize \abs{\U'}=\abs{\V'}}
\sum_{\V'\subset\V}
(-1)^{\abs{\U'}+1}\,(\abs{\U'}+1)!\,\abs{\U'}!\,
\left(\prod_{u\in\U'}\frac{\eps_u}{\tilde u-u}\right)
\left(\prod_{v\in\V'}\frac{\phi_v}{\tilde v-v}\right).
\label{eq:coeff-L}
\eeq

Now we return to the expression \Ref{eq:Lteta-uv-1} for the relevant part of
$(1/2\pi i)\int\l(z)\teta(z)dz$ and
replace the sets $\U$ by $\tilde\U$ and $\V$ by $\tilde\V$.
In the resulting expansion
\beq
\sum_{\scriptsize\U'\subset\tilde\U\atop\scriptsize \abs{\U'}=\abs{\V'}}
\sum_{\scriptsize\V'\subset\tilde\V\atop\scriptsize \V'\neq\emptyset}
(-1)^{\abs{\U'}}\,\abs{\U'}!\,(\abs{\U'}-1)!\,
\left(\prod_{u\in\U'}\eps_u\right)
\left(\prod_{v\in\V'}\phi_v\right)
\L(\U'\cup\V')
\label{eq:Lteta-uv}
\eeq
four types of terms can be distinguished.
%\add{find where the symmetry is lost!}

1. The terms containing no $\tilde u$, $\tilde v$
reproduce obviously the unperturbed expansion \Ref{eq:Lteta-uv-1}.
%Exponentiation of \Ref{eq:Lteta-uv-1} produces the expression
%\Ref{eq:L-repr-norm} for $\pair\U\V$.

2. The terms containing $\tilde u$ but not $\tilde v$.

3. The terms containing $\tilde v$ but not $\tilde u$.

4. The terms containing $\tilde u\in\U'=\U''\cup\{\tilde u\}$ and
$\tilde v\in\V'=\V''\cup\{\tilde v\}$
\beq
\eps_{\tilde u}\phi_{\tilde v}
\sum_{\scriptsize\U''\subset\U\atop\scriptsize \abs{\U''}=\abs{\V''}}
\sum_{\V''\subset\V}
(-1)^{\abs{\U''}+1}\,(\abs{\U''}+1)!\,\abs{\U''}!\,
\left(\prod_{u\in\U''}\eps_u\right)
\left(\prod_{v\in\V''}\phi_v\right)
\L(\U''\cup\V''\cup\{\tilde u,\tilde v\})
\label{eq:Lteta-uv-2}
\eeq
(note that here $\U''$ and $\V''$ are allowed to be empty.

Obviously, only the terms of type 4 produce
$\L(\{\tilde u,\tilde v\})$ via the expansion \Ref{eq:expand-L}.
Noticing that the coefficient
at $\L(\{\tilde u,\tilde v\})$ in the expansion \Ref{eq:expand-L}
of $\L(\U'\cup\V'\cup\{\tilde u,\tilde v\})$
equals
\be
 \prod_{u\in\tilde\U}(\tilde u-u)^{-1}\prod_{v\in\V'}(\tilde v-v)^{-1}
\ee
and
exponentiating \Ref{eq:Lteta-uv-2} we obtain that the coefficient at
$\eps_{\tilde u}\,\phi_{\tilde v}\,\L(\{\tilde u,\tilde v\})$
is given exactly by the expression \Ref{eq:coeff-L} which proves
the theorem \ref{th:charact-FN}.

%\add{finish the proof! explain the sign!}

The recurrence relation allows to give for $\CC(\U,\V;\ell)$ a completely
different representation.

\begin{theo} (Richardson)
Let us fix an ordering of the sets $\U=\{u_1,\ldots,u_N\}$ and
$\V=\{v_1,\ldots,v_N\}$.
The function $\CC(\U,\V;\ell)$  can be expressed as the sum
of $N!$ determinants
\beq
 (-1)^N\CC(\U,\V;\ell)=\sum_{\sigma\in S_N}\det\M^\sigma
\label{eq:Richard}
\eeq
where the sum is taken over all permutations $\sigma$ of indices
$\{1,\ldots,N\}$, and the matrices $\M^\sigma$ are defined as
\beq
 \M^\sigma_{jj}:=\ell(u_j,v_{\sigma_j})+2\sum_{j'\neq j}
(u_j-u_{j'})^{-1}(v_{\sigma_j}-v_{\sigma_{j'}})^{-1}
\eeq

\beq
 \M^\sigma_{jk}:=-2(u_j-u_k)^{-1}(v_{\sigma_j}-v_{\sigma_k})^{-1},
\qquad j\neq k
\eeq
\label{th:sum-of-dets}
\end{theo}

{\bf Proof.} It is easy to verify (see \cite{Ri65}) that the right-hand-side
of (\ref{eq:Richard}) satisfies the same conditions 1--4 of the Theorem 1
which determine uniquely $\CC(\U,\V;\ell)$.
   The theorem \ref{th:norm} is obtained now as a simple corollary.
Indeed, if $\U=\V$ and the parameters $v_k$
satisfy Bethe equations (\ref{eq:bethe}) then in the sum (\ref{eq:Richard})
only the term corresponding to the identical permutation is non-zero,
and the formula for the norm of Bethe function simplifies to \Ref{eq:norm}.
For the details see again \cite{Ri65}.

For a completely different proof of the formula \Ref{eq:norm} based on the
semiclassical expansion of the solutions to the Knizhnik-Zamolodchikov
equation, see \cite{RV}.

%%%%%%%%%%%%%%%%%%%%%%%%%%%%%%%%%%%%%%%%%%%%%%%%%%%%%%%%%%%%%%%%%%%%%%%%%%
\newsection{Discussion}

On a simple example of the Gaudin model we have demonstrated that the
calculation of polynomial correlators simplifies drastically when one
makes use of exponential generating functions. An intriguing and so far
open question is if a similar simplification is possible in case
of XXX or XXZ magnetic chains, when the Lie algebra \Ref{eq:comEFH} is
replaced by the yangian ${\cal Y}[sl_2]$ or, respectively, quantum Lie
algebra ${\cal U}_q[\widehat{sl_2}]$. It is natural to expect that
the complicated formulas for polynomial correlators due to Korepin
\cite{KBI,Kor82} can be somehow simplified by appropriate choice of
generating functions. It would be also interesting to find in this case
an analog of our $\L$-representation and to find an interpretation of
Korepin's `dual fields' in terms of generating functions.

As a simple introductory exercise, one can consider the finite-dimensional
quantum
Lie algebra ${\cal U}_q[sl_2]$ and reproduce the derivation given in
the beginning of the section 4.
The necessary formulas concerning the Gauss decomposition for
${\cal U}_q[sl_2]$ can be found in \cite{MV94}. The main new feature in
comparison to the Lie algebraic case is that the parameters of expansion
$\eps$, $\eta$ and $\phi$ become noncommuting.

%%%%%%%%%%%%%%%%%%%%%%%%%%%%%%%%%%%%%%%%%%%%%%%%%%%%%%%%%%%%%%%%%%%%%%%%%

\end{document}